\documentclass[preprint,12pt]{elsarticle}

\usepackage{graphicx}

\usepackage{amsmath}
\usepackage[latin1]{inputenc}
\usepackage{txfonts}
\usepackage{hyperref}

\journal{Physica A}

\begin{document}

\begin{frontmatter}

\title{Synchrony-optimized networks of \\ Kuramoto oscillators with inertia}
\author{Rafael S. Pinto}\ead{rsoaresp@gmail.com}

\address{Instituto de F\'\i sica ``Gleb Wataghin'', UNICAMP, 13083-859 Campinas, SP, Brazil.}

\author{Alberto Saa}\ead{asaa@ime.unicamp.br}

\address{
Departamento de Matem\'atica Aplicada, 
 UNICAMP,  13083-859 Campinas, SP, Brazil.}

\begin{abstract}
We investigate synchronization in networks of Kuramoto oscillators with inertia. 
  More specifically, we introduce a rewiring algorithm consisting basically in a  {\em hill climb} scheme in which the edges
of the network are swapped in order to enhance its    synchronization capacity.
 We show that the 
the synchrony-optimized networks generated by our algorithm have some interesting topological and dynamical properties. In particular, they typically
exhibit an anticipation of the synchronization onset and are more robust against certain types of perturbations. 
We consider  synthetic random networks and also a network with a topology based in an approximated model of the (high voltage) power grid of Spain, since
networks of Kuramoto oscillators with inertia have been used recently as simplified models for
power grids, for which synchronization is obviously a crucial issue. Despite the  extreme simplifications adopted in these models, our results, among others recently obtained in the literature, may provide interesting principles to guide the future growth and development of real-world
  grids, specially 
in the case of a change of the current paradigm of centralized towards distributed generation power grids.
\end{abstract}

\begin{keyword}
\PACS 89.75.Fb \sep 05.45.Xt \sep 89.75.-k
\end{keyword}

\end{frontmatter}

\section{Introduction}

Since Thomas Edison's Pearl Street Station in Manhattan started operating in 1882, the power grids have continued to grow and are today probably the largest machines ever built 
\cite{strogatz2003,kundur1994}. Their growth is still far away from being complete, since the pursuit of renewable sources of energy and new technologies drive the integration of different power grids into continental  machines as, for instance, the paradigmatic case of Western Europe.
(For an approximated description of the   western european interconnected high voltage power grid, see 
\cite{Bialek,hutcheon}.)
 The widespread use of alternating current   creates the necessity of keeping the whole power grid synchronized, and
  a disruption in this synchronization may cause malfunctioning, leading to power outages with possible catastrophic proportions in real scenarios.

The phenomenon of synchronization has been  studied for a long time \cite{strogatz2000,pikovsky2001,arenas2008,acebron2005} in different areas of knowledge. It is present in a myriad of situations, arising naturally in many areas of biology, physics, social sciences, etc. However, It has been
only recently that a  \emph{complex system} approach has been devised to study the synchronization of power grids \cite{filatrella2008} (See also, for a recent review, \cite{NishMotter}). Typically, a single power plant is a complicated machine, with  a lot of tunable  parameters necessary to its correct functioning. Although power grids can be, and surely are, analyzed and studied in all their finer details, taking into account hundreds of power plants, substations, transmission lines, and many other devices, the idea  here   is to focus on the complexity of the underlying network of connections \cite{newman2003,newman2010} and its role on the overall synchronization process. In order to achieve such a goal, one treats the power plants as simple generators and the loads on the other side of transmission lines as passive machines. Energy balance in this context yields a set of equations   known as the Kuramoto Model with inertia \cite{acebron2005}. Recently, we have witnessed  many works devoted to the analysis of power grids in this context of complex system as, for instance, the analysis of the European power grid \cite{lozano2012}, the effects of decentralization of energy production in the British power grid \cite{rohden2012}, the identification of parameters in individual vertices that turn the synchronous state more stable \cite{motter2013}, the existence of Braess's paradox \cite{witthaut2012,witthaut2013}, the role of the   topology \cite{rohden2014,PNAS,decentral} and assortative mixing \cite{assort} on the network synchronization 
and control, and a stability analysis of blackouts using   basin-stability   measures \cite{menck2014}.

In this paper, we study the optimization, in order to favor synchronization, of   networks of Kuramoto oscillators with inertia 
modelling   power grids. More specifically, we adapt a   algorithm previously proposed in 
\cite{brede2008a,brede2008b} to optimize the synchronization capacity of a network built from usual Kuramoto oscillators to the case of 
oscillators with inertia.   We study the main topological and dynamical properties of the 
synchrony-optimized networks
generated by our algorithm 
  and their   robustness for edges (corresponding to transmission lines) removal and other perturbations 
  which could mimic consumption peaks (or generation shortages) in real power grids.
  Our results show that the optimized networks tend to be more robust against 
the perturbations mimicking   
  consumption peaks for a wider  range of parameters 
when compared to   the non-optimal networks, and no differences were detected with respect  to edge removals.
 
 The paper is organized as follows. In Sec. II, we review briefly  the power grid model based on Kuramoto oscillators with
inertia \cite{filatrella2008} and discuss some of its properties with relevance to our analysis.  Sec. III is devoted to the introduction of our optimization algorithm. In Sec. IV, we show the numerical results obtained for   synthetic random networks   and also for a network topologically
based in an approximation of the Spanish high voltage power grid.

\section{Kuramoto oscillators with inertia}

For the sake of completeness, we will briefly review here the basic   equations   for the 
Kuramoto oscillators with inertia used in Ref. \cite{filatrella2008} for a simplified description of power grids. Our main goal is to  derive some simple results concerning the synchronization  of the underlying network which are important to our analysis. In this simplified description, 
a power grid  is represented   by an undirected  network  composed of $N=N_{G}+N_{C}$ vertices corresponding to two types of machines: $N_{G}$ generators and $N_{C}$ consumers (motors), which do not need to be equal in number necessarily.  The power transmission lines correspond  to the $m$ edges connecting the vertices. The connectivity pattern is described by the usual symmetric adjacency matrix $A$, with elements $a_{ij}$ such that $a_{ij} = 1$ if vertices $i$ and $j$ are connected, and $a_{ij} = 0$ otherwise.

Each individual element $i$ of the network corresponds to a synchronous machine, generator or consumer, characterized by a power $\tilde{P}_{i}$, which  is positive for generators and negative for the consumers. For each network vertex, simple energy balance  implies that this power must be equal to the sum of three contributions: the rate of change of the machine kinetic energy 
\begin{equation}
P^{\rm kin}_{i} = I_{i} \ddot{\theta}_{i}\dot{\theta}_{i}
\end{equation}
where $I_i$ and $\theta_i$ stand for, respectively, the moment of inertia and the phase of the $i$-th generator/consumer;  the rate which energy is dissipated trough friction
\begin{equation}
 P^{\rm diss}_{i} = \gamma_i \dot{\theta}_{i}^{2}, 
\end{equation}
where $\gamma_i$ is the dissipation constant associated with the machine at vertex $i$; and the total power transmitted to other vertices. In particular,  the power transmitted from vertex $i$ to $j$ is given by 
 \begin{equation}
  P^{\rm trans}_{ij} = - P_{ij}^{\rm max} \sin(\theta_j - \theta_i) ,
 \end{equation}
   where $P_{ij}^{\rm max}$ represent the maximum power that can be transmitted  along the
   transmission line connecting $i$ and $j$   vertices. Summing all   terms, one has
\begin{equation}
\tilde{P}_{i} = I_i \ddot{\theta}_{i} \dot{\theta}_{i} + \gamma_i \dot{\theta}_{i}^{2} - \sum_{j=1}^{N}   P_{ij}^{\rm max} \sin(\theta_j - \theta_i).
\label{energy_conservation}
\end{equation}
From now on, we restrict ourselves to the idealization often assumed   for power grids in this context, namely that
all elements in the grid have the same moment of inertia $I$ and the same dissipation constant $\gamma$, and that 
all transmission lines have the same capacity of transmission $P^{\rm max}$.
Relaxing these hypotheses does not, apparently, lead  to new interesting   dynamical behaviors, but   makes   the whole analysis
much more intricate. For the proper functioning of the power grid, all of the elements  must operate with the same frequency $\Omega$ (for instance, $50$ or $60$ Hz for real power grids). In order to take into account small fluctuations around this value, we write the element phases as
\begin{equation}
\label{Omega}
  \theta_{i}(t) = \Omega t + \phi_{i}(t) ,
\end{equation}  
with $|\dot\phi_i|\ll \Omega$. Taking into account the above simplifications 
and keeping only linear terms in the perturbation 
  $\dot\phi_{i}(t)$ in  (\ref{energy_conservation}), one has the so-called Kuramoto  equations with inertia
\begin{equation}
\frac{d^2\phi_i}{dt^2} = P_i -\alpha \frac{d\phi_i}{dt} + K\sum_{j=1}^{N} a_{ij} \sin (\phi_j - \phi_i),
\label{kuramoto_power_grid_definition}
\end{equation}
where $P_i = \left( \tilde{P}_i - \gamma \Omega^2 \right) / I\Omega$, $\alpha = 2\gamma/I$, $K = P^{\rm max}/I\Omega$,
and $a_{ij}$ stands for the usual adjacency matrix for the underlying network. The natural timescale for this system is
$\alpha^{-1}$, and the parameters  $P_i$ and $K$ have both the dimension of $\alpha^{2}$.
Some  useful information can be obtained even before solving 
 equations  (\ref{kuramoto_power_grid_definition}). The first point to notice is the existence of a synchronized stationary state with the grid frequency $\Omega$. For such a  state, one needs $\ddot{\phi}_i = \dot{\phi}_i = 0$, which means that the stationary phases $\phi_i$ of the oscillators satisfy the equation
\begin{equation}
P_i + K\sum_{j=1}^{N} a_{ij} \sin (\phi_j - \phi_i) = 0,
\label{stationary_phases}
\end{equation}
where, incidentally, the timescale $\alpha^{-1}$ plays no role.  Summing both sides of (\ref{stationary_phases}) with respect to the index $i$ and using the fact that adjacency matrix $a_{ij}$ of an undirected network is symmetric, we have
\begin{equation}
\sum_iP_i=0,
\label{sum}
\end{equation}
since the sine function is odd.
 In this way, a necessary condition for the existence of the synchronized stationary state with frequency $\Omega$ is that the 
individual machine  
 powers $\tilde P_i$ obey
\begin{eqnarray}
\sum_i\tilde P_i= N\gamma\Omega^2,
\end{eqnarray}
which obviously is nothing else than a statement of energy conservation for the whole network (neglecting, of course, transmission losses). Incidentally, the relaxing of   condition (\ref{sum}) can be considered as situations with temporary  peaks of
consumption or shortage of energy generation, which  will be considered later since they present
interesting dynamical behavior.

In order to calculate the network average frequency perturbation 
\begin{equation}
\langle \dot{\phi} \rangle =  \frac{1}{N}\sum_{i=1}^{N} \frac{d\phi_i}{dt}.
\end{equation}
we sum both sides of equation (\ref{kuramoto_power_grid_definition})  and use the  symmetry of  $a_{ij}$, resulting in 
\begin{equation}
\label{triv}
\frac{d}{dt} \langle \dot{\phi} \rangle = 
  - \alpha  \langle \dot{\phi} \rangle,
\end{equation}
where (\ref{sum}) was used. 
Equation (\ref{triv}) can be trivially solved, leading  to 
\begin{equation}
\langle \phi \rangle   = \langle \phi_0 \rangle + \frac{1}{\alpha} \langle \dot{\phi}_0 \rangle  \left(1 - e^{-\alpha t}\right),
\label{mean_phase}
\end{equation}
The average frequency perturbation 
vanishes in the limit $t \rightarrow \infty$,   but this does not imply that each frequency $\dot\phi_i$ converges to zero,
since they can also attain a state with symmetric distribution of positive and negative frequencies with null mean. However, if we have synchronization, they do vanish individually.  Equations (\ref{mean_phase}) confirm an earlier numerical result that was verified in \cite{rohden2014}: the time scale to reach asymptotically stationary states in this kind of network is proportional to $\alpha^{-1}$. 
Notice also that  the asymptotic value of the average phase depends on the initial conditions and 
also on the parameter $\alpha$.

In order to analyze the synchronization process in our networks, we will use the order parameter $z(t)$ introduced originally by Kuramoto,
\begin{equation}
z(t) = r(t)e^{i\psi(t)} = \frac{1}{N}\sum_{j=1}^{N}e^{i\phi_{j}(t)},
\label{kuramoto_order_parameter_definition}
\end{equation}
which corresponds to the centroid of the phases if they are viewed as a swarm of points moving around the unit circle. For incoherent motion, the phases are scattered on the circle
homogeneously and $r \approx N^{-1/2}$ for large N  as a consequence of the central limit theorem, while for a synchronized state the
points should move in a single lump and, consequently, $r \approx 1$. Since our equations are of second order, we will also use the mean squared
(dimensionless) frequency,
\begin{equation}
v^{2}(t) = \frac{1}{\alpha^2}\left\langle \dot{\phi_j}^{2} \right\rangle = \frac{1}{\alpha^2 N} \sum_{j=1}^{N} \dot{\phi_j}^{2}(t).
\label{velocity_order_parameter_definition}
\end{equation}
Synchronization for the power grid requires, of course, that all elements have the common frequency $\Omega$, which
implies that $v^2(t) = 0$. In our calculations, we will use extensively the averages
\begin{equation}
r =  \frac{1}{\delta T} \int_{T }^{T +\delta T}\left| z(t) \right| \, dt ,
\end{equation}
and
\begin{equation}
v^2 = \frac{1}{\delta T} \int_{T}^{T +\delta T} v^2(t)\, dt,
\end{equation}
where value of $T $ must be long enough to guarantee that a  stationary state has been reached,
and  $\delta T$ cannot be too small in order to assure good statistics.  

As already mentioned,  power grids can be  analyzed and studied in great detail, taking into account hundreds of different 
power plants, substations, transmission lines, and many other devices. 
On the other hand, the idea used   here   is to focus on the complexity of the underlying network of connections, as done, for instance,
in \cite{filatrella2008,NishMotter,newman2003,newman2010}, and its role on the overall synchronization process. 
The universality of the dynamical behavior of synchronization in complex networks implies that the simplified description of a 
real power grid by means of a network of Kuramoto oscillators with inertia is more than a formal analogy, and it should be capable
of providing  a good 
qualitative description of the synchronization process in large networks. However, it is clear
that the simplified model will not be able to describe satisfactorily many realistic dynamical properties of power grids due
to the   rough approximations used here.

\section{The Optimization algorithm}

The optimization algorithm employed here is adapted from that one introduced in \cite{brede2008a,brede2008b} for the original Kuramoto model, which corresponds to the model of the last section without inertia ($I=0$). Here, by optimization, we mean a rewiring of the edges such that the new network has higher values of the   order parameter $r$, corresponding, of course, to a situation of enhanced synchronization.
As it occurs typically in synchronization problems, there is a phase transition from a non synchronized state  to a synchronized one occurring at a critical value of the parameter $K$, {\em i.e.}, for $K > K_c$, the order parameter $r$ is an increasing function of $K$, while for $K<K_c$ we have typically $r\approx 0$ (no synchronization). 
The strategy of the algorithm is roughly the following. We choose a value of $K^{*} > K_c$ such that $r(K^{*})$ is reasonably high. One then considers a rewiring: a randomly selected edge connecting two vertices is removed if it does not disconnect the network, and  two randomly chosen disconnected vertices are connected; and the new value of $r(K^{*})$ is evaluated.
If the rewiring results in a higher value for $r(K^{*})$, one keeps the modification or, otherwise, one discharges it and returns the network to its previous configuration. This procedure is repeated until  $r(K^{*})$ attains a maximum value. In practice, our algorithm limits the maximum number of iterations   and also stop after a certain number of consecutive iterations failed to achieve a higher value of $r$. These edge swaps preserve  the average degree of the initial network (as the number of edges is kept the same), but not the degree distribution. Nevertheless, we can easily include in the optimization process other constraints such as, for instance, accepting a rewiring only if the degrees of the vertices or the length of the transmission lines are lower than a pre-determined value. Here, however, we consider only the simplest case of improving the synchronization properties by only limiting the number of and edges (transmission lines). For a discussion on possible rewiring constraints for realistic power lines, see 
\cite{schneider2011,Louzada2013}.

As for the original Kuramoto model \cite{brede2008a,brede2008b}, an earlier onset of synchronization is observed for the optimal networks.
In other words, synchrony-optimized networks typically have  smaller $K_c$.
  Moreover, the properties of the optimal network are independent from the initial conditions and also from the precise value of $K^*$ for which the rewiring is done, as long as it is chosen to satisfy $K^{*} > K_c$, {\em i.e.}, it indeed corresponds to a synchronization regime for
  the original network.
Our {hill climb} algorithm produces networks   exhibiting  the desired properties of higher values of the order parameter $r$ and smaller values of $v^2$ for all the values of $K > K_c$, and not just for the value  $K=K^{*}$ for which the optimization was performed. 
For small networks  (typically with $N$ up to 20), the optimal network seems to be unique, since all runs return the very same network configuration. However, for larger networks, different runs of our algorithm can lead to different optimized networks, but their properties are essentially the same, showing just some small fluctuations over the average values. In this way, strictly speaking, we cannot guarantee that the results found with the algorithm are global maxima, and this is particularly important for large networks.  Nevertheless, the returned optimal networks 
always show a substantial improvement of the synchronization properties when compared with the original ones.

\section{Results}

In this section, we will show some of the results obtained using our optimization algorithm. We analyze two situations  which we call decentralized and centralized energy production. The first case corresponds to a network with $N$ vertices in which half of them are generators   and the other half corresponds to consumers. This situation tries to mimic future development in power grids where many (perhaps small) power plants, with different energy sources, will be connected to the grid. The second case, on the other hand, represent the current situation, with energy production restricted  to a small number (compared to the number of consumers) of large power plants.  
Our numerical simulations were done by using the SciPy package for python \cite{SciPy}. The system of ordinary differential equations (\ref{kuramoto_power_grid_definition}), in
particular, is solved with SciPy {\tt odeint} routine, which is indeed a implementation of {\tt lsoda} from the FORTRAN library odepack.
We have also made extensive use of the NetworkX package\cite{NetworkX} for 
calculating network properties and characteristic parameters and for creating the   network graphs.

\subsection{Decentralized power grids}

\begin{figure}[t]
\includegraphics[scale=0.32]{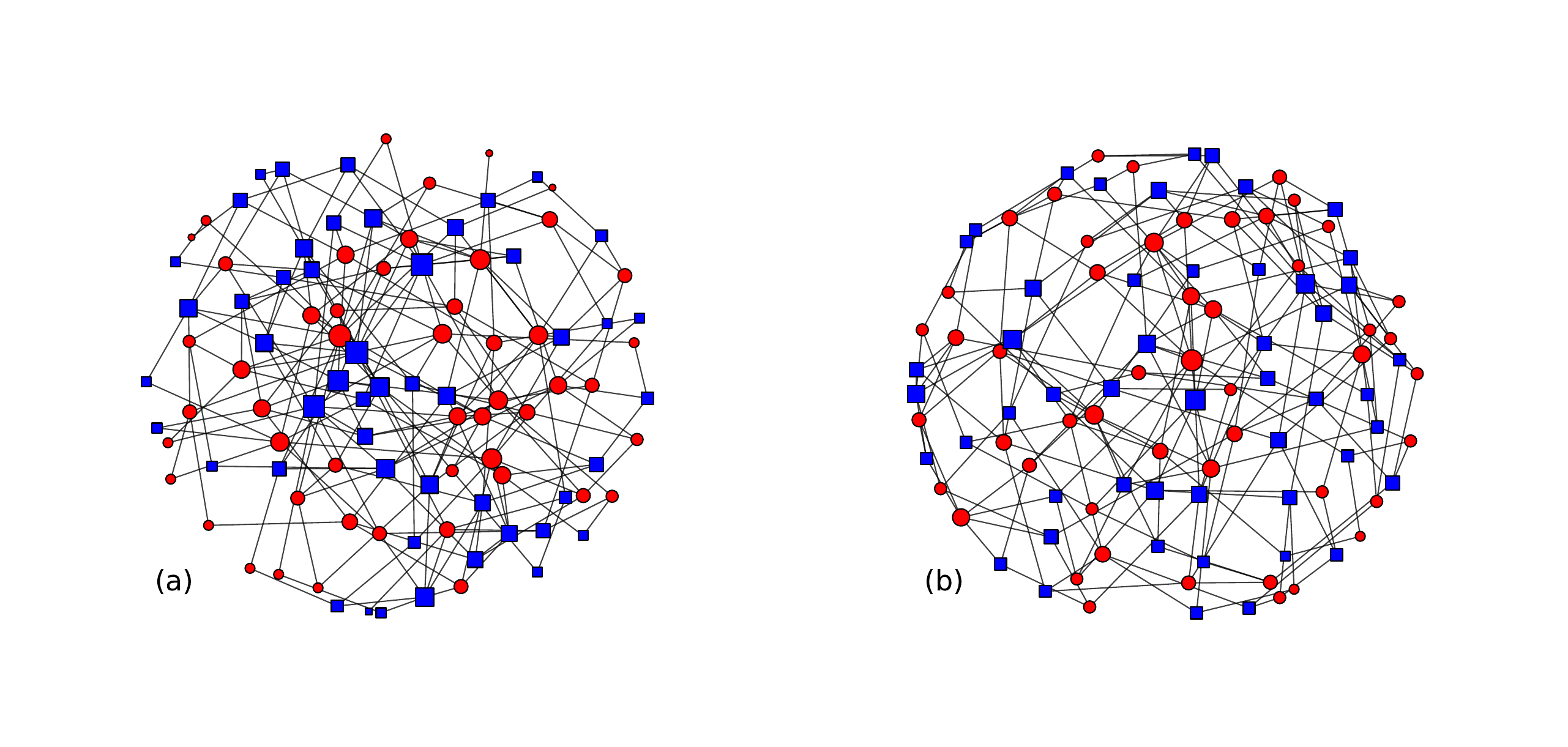} 
\caption{An example of optimized network (figure b) obtained with the \emph{hill climb} algorithm discussed in the text. We used an Erdos-Renyi network (figure a) with $n = 100$ vertices and $m = 202$ edges as initial condition. There are 50 blue squares and 50 red circles representing consumers and generators, respectively. Vertex's size is proportional to its degree.}
\label{optimization_from_er}
\end{figure}

Let us illustrate the case of  decentralized  power grids  by considering  (synthetic) networks built  from the Erdos-Renyi (ER) model \cite{newman2010} with $n = 100$ vertices and average degree $\langle k \rangle = 4$. In ER networks, edges are added with probability $p = \langle k \rangle / n$, independently of the   vertex degree. Half of the vertices were randomly selected as generators  with $P = 1\alpha^2$, and the other half as consumers, with $P = -1\alpha^2$.  
An example  of the network considered in the optimization process is depicted in Figure \ref{optimization_from_er}.
 The left panel shows the original power grid, whereas the right one shows   the optimized network  obtained by employing the algorithm.
We considered  an ensemble of 12 ER random networks, again with $n = 100$ and $\langle k \rangle = 4$, carefully chosen to have the same number of edges. The role  of each vertex (generator or consumer) is the same for all networks in the ensemble. 

\begin{figure}[ht]
\centering
\includegraphics[scale=0.32]{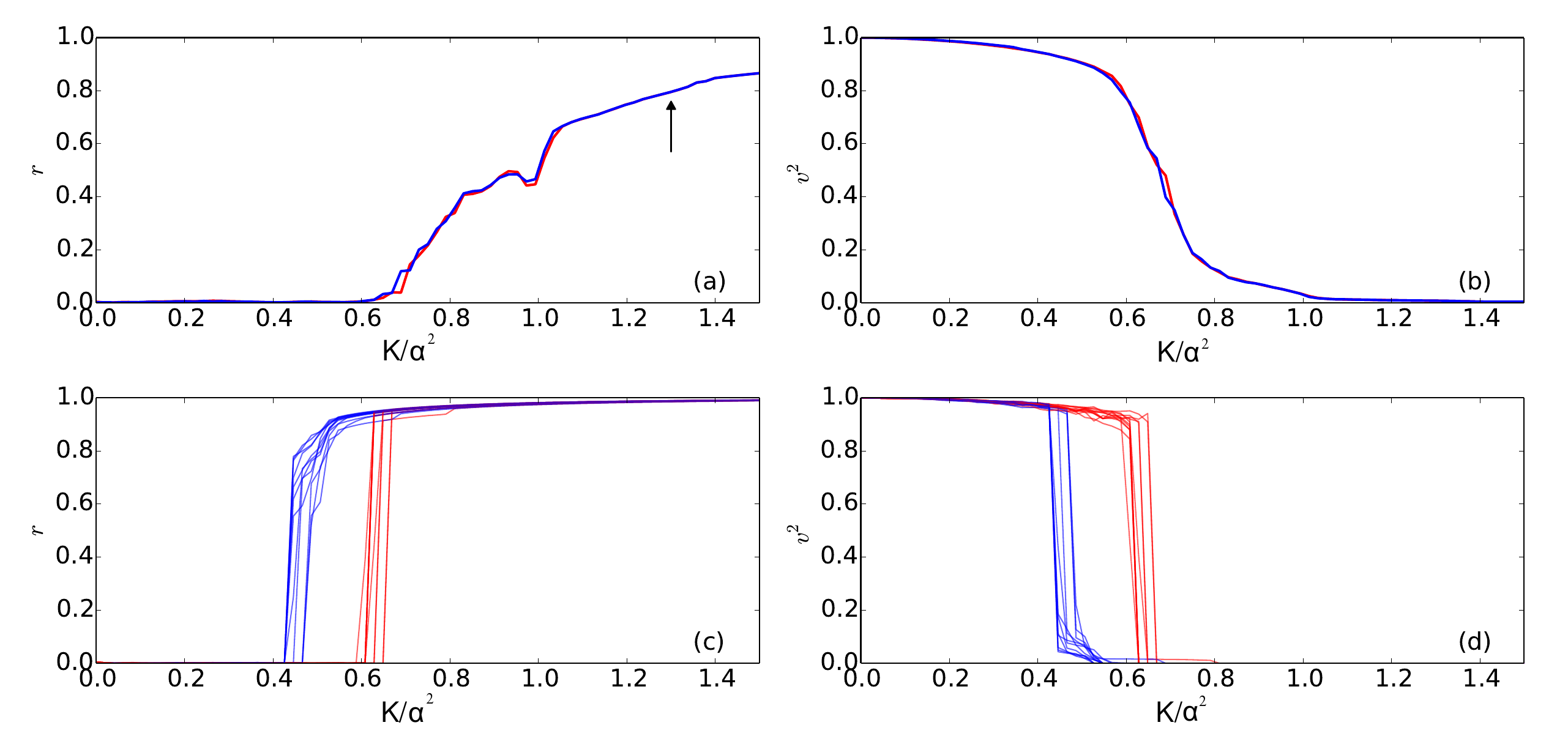}
\caption{Synchronization diagrams for the power grids. Both the forward (red) and backward (blue) continuations in $K$, with $\delta K = 0.02\alpha^2$, are shown. Panels (a) and (b) show, respectively, the order parameter $r$  and the mean squared frequency $v^2$    over the ensemble of 12 ER networks, suggesting a second order phase transition, since the forward and backward continuations roughly coincide. In turn, panels (c) and (d) show  the results of the optimization algorithm performed on each network. In this case, the phase transition appears to be of first order with a hysteresis behavior. The arrow in panel (a) shows the value of $K^*$ where the optimization process was performed.  }
\label{er_opt_er_order_par}
\end{figure}

We call a synchronization diagram of a network  the graphics of  the dimensionless parameters $r$ and and $v^2$, given 
respectively by the equations 
(\ref{kuramoto_order_parameter_definition}) and (\ref{velocity_order_parameter_definition}), as functions of
the parameter $K$.  
Figures  \ref{er_opt_er_order_par}a and \ref{er_opt_er_order_par}b depict the synchronization diagrams for the non-optimal 
network ensemble, while \ref{er_opt_er_order_par}c and \ref{er_opt_er_order_par}d correspond to the
synchrony-optimized outputs from the algorithm.
The graphics of Figure \ref{er_opt_er_order_par} were plotted starting from $K = 0$,  with increments of $\delta K = 0.02\alpha^2$ until $K$ reached the value $K = 1.5\alpha^2$. At each step, the outcome of the last run is used  as initial conditions to the next one.
From this point, we reversed the step direction and started to decrease the value of $K$, again with steps of size $\delta K=0.02\alpha^2$. By means of this procedure, we have two synchronization diagrams, called the \emph{forward} and \emph{backward} continuations\cite{gomez-gardenes2011}, respectively, 
and a hysteresis loop is formed in the situations where they do not coincide. 

For the non-optimized networks, both the forward and backward continuations are the same  except for some small fluctuations. The absence
of any hysteresis behavior indicates  a sort of second order phase transition.  For the optimized network, however,   interesting new behaviors arise. Firstly, we have an earlier onset of synchronization (smaller values of $K_c$), and the values of the order parameter $r$ attain   higher values, indicating that the phases, although not all equal, have a much narrower distribution than for the non-optimized case. Secondly, and very interesting, the type of the phase transition seems to change from second to first order with a hysteresis behavior, as the forward and backward curves no longer match each other.
The existence of a first order phase transition with hysteresis behavior was already studied for the model (\ref{kuramoto_power_grid_definition}) in the case of an \emph{all-to-all} topology  \cite{tanaka1997}. (See also \cite{hyster}). It is clear that the topology plays a fundamental role in the synchronization properties of a network, since a rewiring of the network, keeping everything else unchanged, is capable of modifying the kind of synchronization transition. This kind of first-order phase transition for the synchronization is the key dynamical point of the so-called explosive synchronization behavior, which is now under intensive investigation, see  \cite{gomez-gardenes2011} and \cite{explosive}, for instance. 

\begin{figure}[t]
\centering
\includegraphics[scale=0.32]{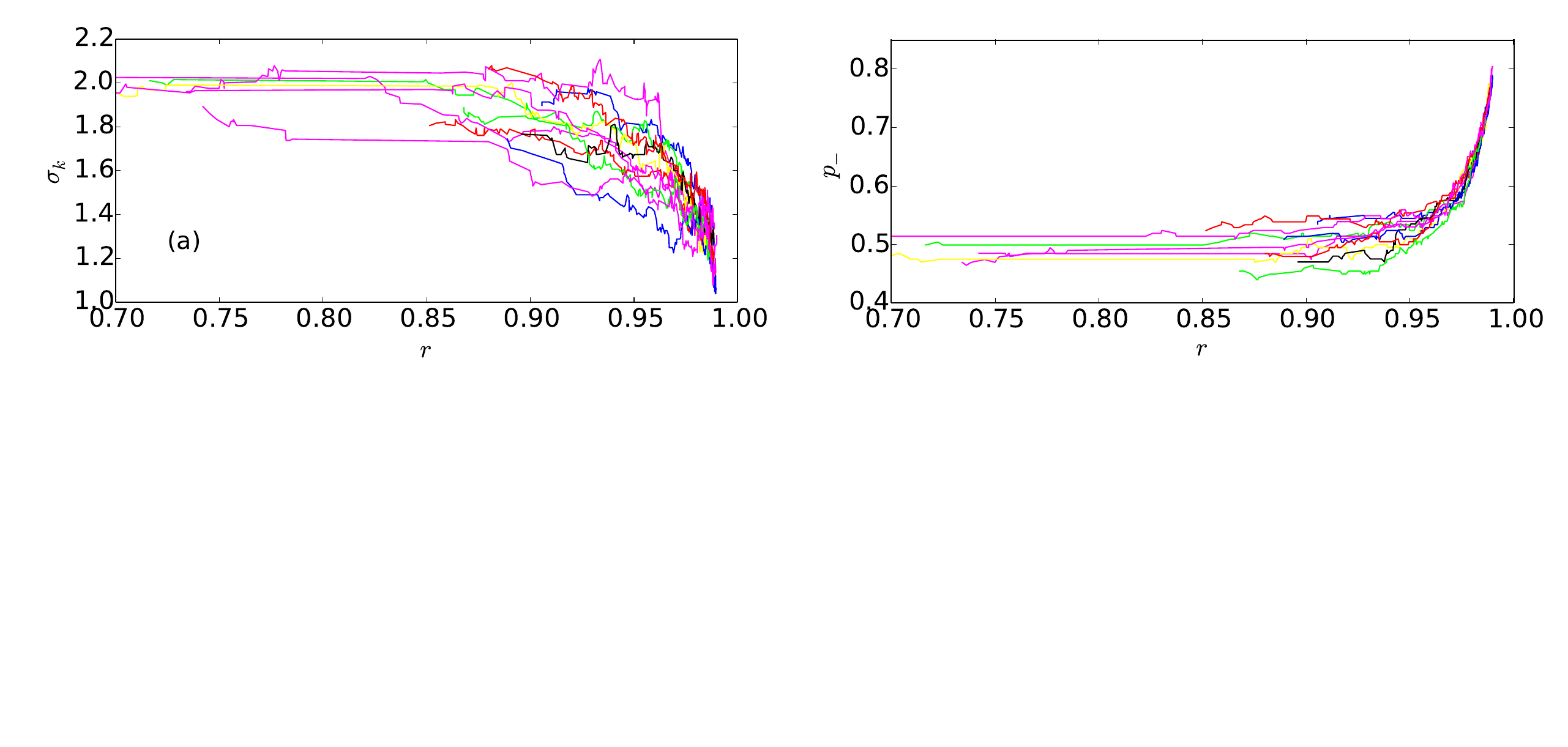}
\caption{The evolution of some network characteristic parameters through the optimization algorithm steps. Each line correspond to a network of the ensemble discussed in Fig. \ref{er_opt_er_order_par}. The panels depict: (a) the standard deviation for the degree distribution $\sigma_k$
 and (b) the fraction $p_{-}$ of edges connecting consumers to generators.}
\label{optimization_properties}
\end{figure}

We have studied also the evolution of the network 
topology through  the optimization process. For each network in the ensemble, we have followed 
the evolution of some topological characteristic parameters at each step of the optimization algorithm. 
The standard deviation for the degree distribution,  $\sigma_k$, and the fraction $p_{-}$ of edges connecting consumers and generators, show
clearly a trend  process, as depicted in Figure \ref{optimization_properties}. 
We can see  that  the network topology becomes more homogeneous at each step of the algorithm, since  $\sigma_k$ decreases while the
average degree of the network is kept constant. 
On the other hand, the fraction $p_{-}$ tends to increase monotonically to values close to 1, in agreement with the
results obtained previously in  \cite{brede2008a}. The synchrony-optimized
networks obtained from our algorithm have always smaller $\sigma_k$ and larger $p_{-}$ when compared with the non-optimal ones. 
Other relevant network topological  parameters such as the average shortest path length $\langle l \rangle$ and the clustering coefficient $C$ 
do not exhibit any preferred trend  throughout  the optimization process and, thus, they do not seem to be relevant to characterize the synchrony-optimized
networks. Accordingly to these results,  the best way to build a power grid from the synchronization point of view
 is to guarantee a homogeneous grid (small $\sigma_k$) in which the transmission lines connect preferably  consumers to generators
 (large $p_{-}$). Knowing in advance this optimal pattern of connections, we could use the algorithm proposed in \cite{kelly2011} for optimizing decentralized power grids. Since it is an algorithm to maximize $p_{-}$ for the network, it may provide synchrony-optimized decentralized power grids as well. The advantage of the algorithm of \cite{kelly2011} is that it is, typically,   faster than ours.

\begin{figure}[t]
\centering
\includegraphics[scale=0.32]{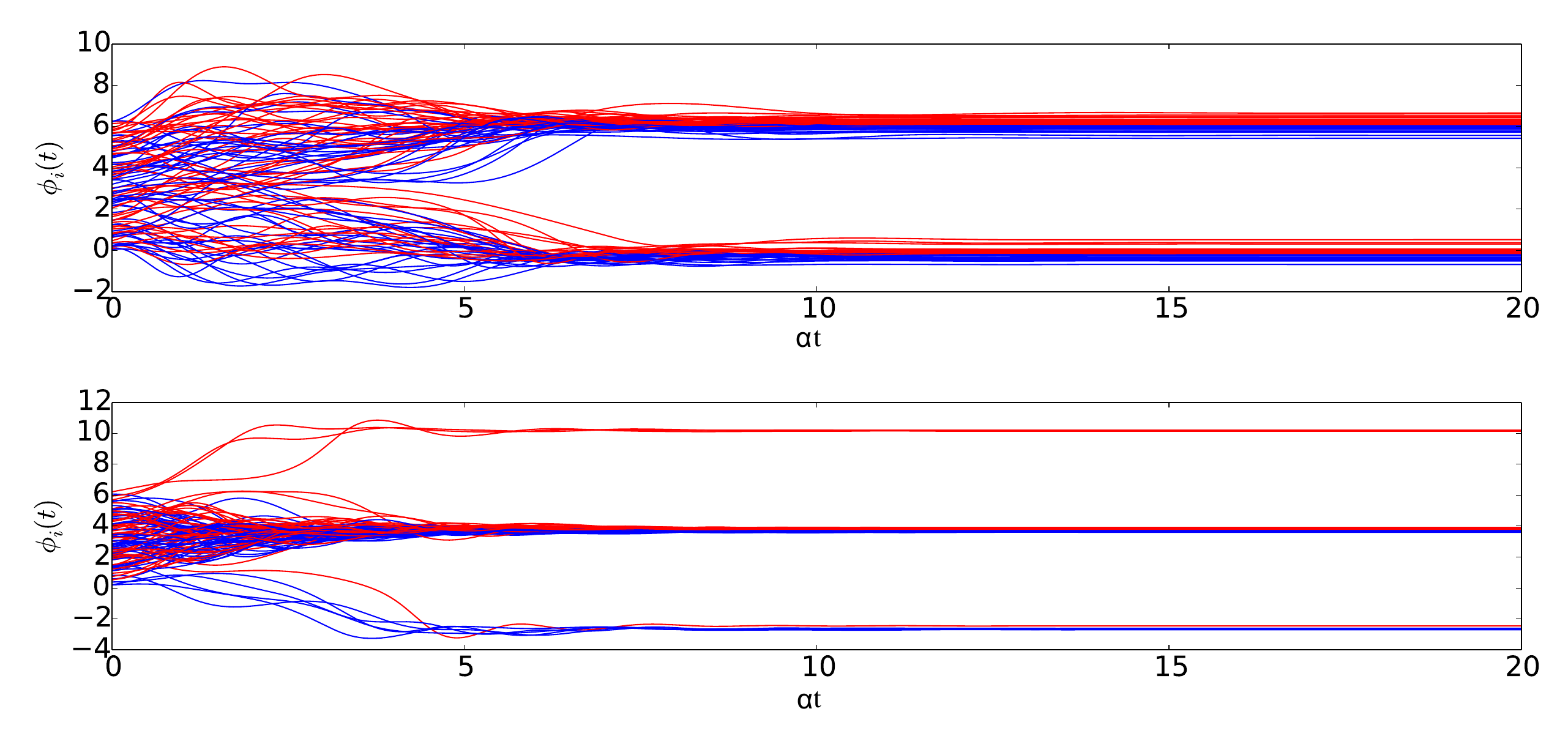}
\caption{The phases of the generators (red lines) and consumers (blue lines) as a function of   time for the networks in figure \ref{optimization_from_er}. Top and bottom panels show the phases for the non-optimized and optimized networks, respectively,
for $K = 2.0\alpha^2$. The initial phases were randomly draw from the uniform distribution in $(0, 2\pi)$ and the frequencies from the uniform distribution in $(0, 1)$.}
\label{phases_er}
\end{figure}

\begin{figure}[t]
\centering
\includegraphics[scale=0.32]{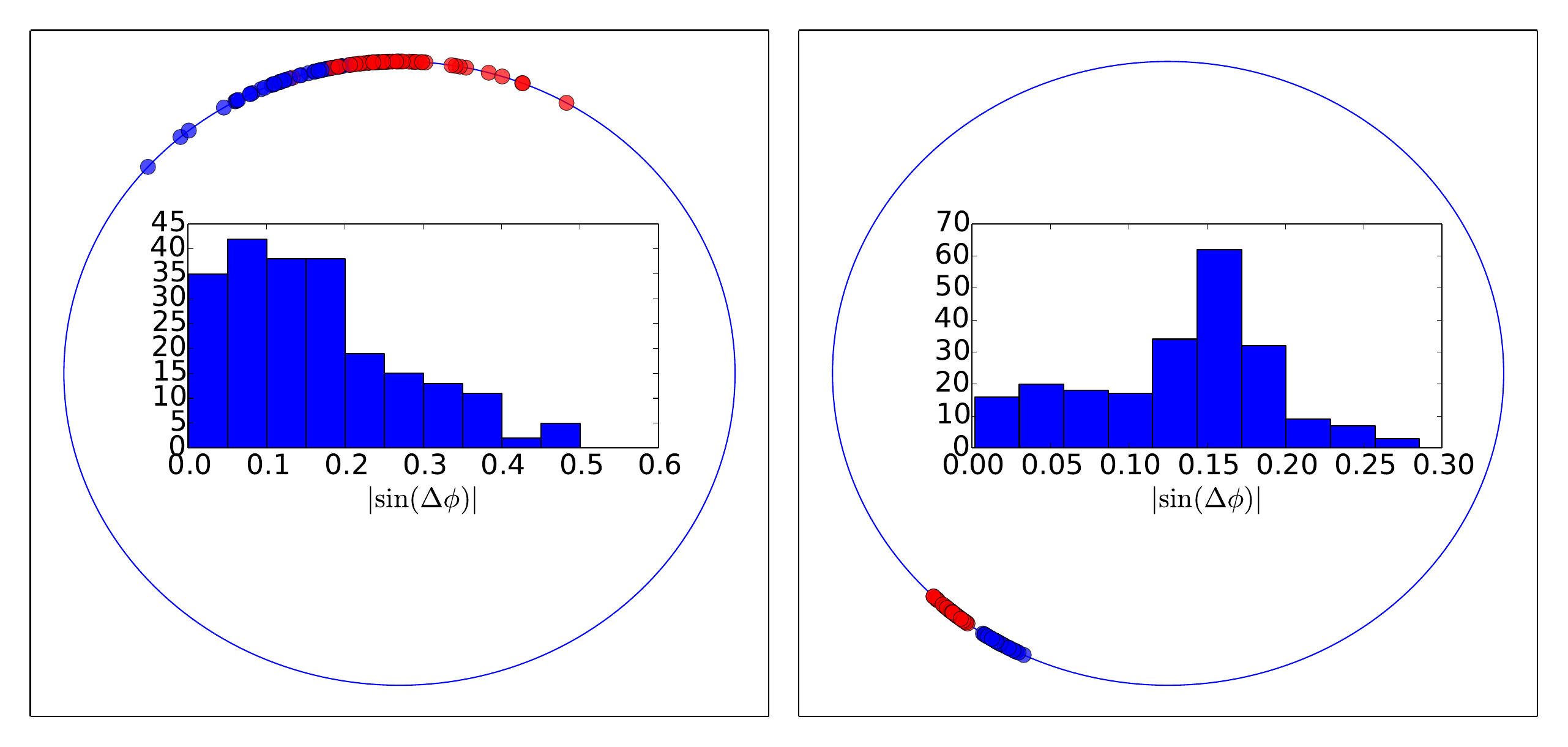}
\caption{The phases of the generators (red circles) and consumers (blue circles) depicted around the unit circle,
at a given time in the stationary regime ($t=20\alpha^{-1}$),  for
the networks of Figure (\ref{phases_er}). Inserted, the histograms of $\left|\sin\left(\phi_i-\phi_j \right) \right|$ for the
 edge connecting
vertices $i$ and $j$. The optimized grid has a narrower distribution of phases between connected vertices.
}
\label{phases_circle}
\end{figure}

Another interesting point is  the individual dynamics of the generators and consumers for the non-optimal and optimal networks.
 As we have seen earlier, the time scale to reach the stationary state is determined only by the dissipation parameter $\alpha$, see equation (\ref{mean_phase}). Nevertheless, the way that the individual  phases distribute themselves is determined by the topology of the power grid, see equation (\ref{stationary_phases}). Figure \ref{phases_er} shows the evolutions of the phases for the networks in Figure \ref{optimization_from_er}.
Interestingly, for the same value of   $K$, we found that for the optimal  power grids the phases   have a much narrower distribution compared to the results of the non-optimized network.
Figure (\ref{phases_circle})
 depicts the phases of Figure (\ref{phases_er}) around the unitary circle in the stationary regime,
and the corresponding histogram of $\left|\sin\left(\phi_i-\phi_j \right) \right|$ for the   edge connecting
vertices $i$ and $j$. As we see, the optimized grids have typically   narrower phase differences  between connected vertices   and,   consequently,
    power is transmitted more homogeneously in the network.
This has important implications for the power grid functioning. The phase difference between connected vertices is obviously related to the transmitted power, but also to the losses in the transmission lines \cite{watanabe2013}. The ability of keeping the phases closer to each other with the same value of  $K$ has a important role in the search for efficiency.

We have also studied the robustness of the networks against edge removal. The size of the giant component $S(m)$ is a common topological measure employed in the analysis of how connected  a network  remains after removing $m$ edges \cite{zeng2012}. 
 We have considered three different edge removals rules: random removal; the so-called degree product rule, 
 where one removes first the edges with the highest product of the degrees of the vertices connected to them; and the so-called edge-betweenness rule, where the edges with highest edge-betweenness are removed firstly.  Our analyses 
  show that the behavior of optimal networks  with respect to edge removal is much similar to the non-optimal case,  for the three edge removals rules considered. In order to increase the robustness for edge removal, it is possible also to include in the optimization process a measure of robustness, as the ones proposed in \cite{schneider2011} and \cite{zeng2012}, for instance. It is straightforward to change the algorithm in such a way  that an edge swap is accepted only if it increases both the synchronization order parameter and the adopted robustness measure. Nevertheless, our results
shows   
   that our optimization procedure does not weaken the robustness of the original network against edge removals.
   
Finally, we stress that all the explicit cases discussed in this section have been chosen as typical illustrations for the algorithm
process. We have made exhaustive numerical studies, and our conclusions do not depend on either the number $N$ of vertices, or the
average degree $\langle k\rangle$ of the ER network.  In fact, similar results also hold for other type of synthetic networks as,
for instance,    those ones   generated from the Barabasi-Albert model \cite{barabasi1999}, which is based in a preferential attachment mechanism 
leading to   power law degree distributions. 
The main properties of the optimal networks   are fairly the same, corroborating  the results of \cite{brede2008a}  
suggesting 
that the synchrony-optimized  networks are independent of the initial conditions (networks), once we keep   fixed the set of powers $P_i$.

\subsection{Centralized power grids}

\begin{figure}[t]
\centering
\includegraphics[scale=0.32]{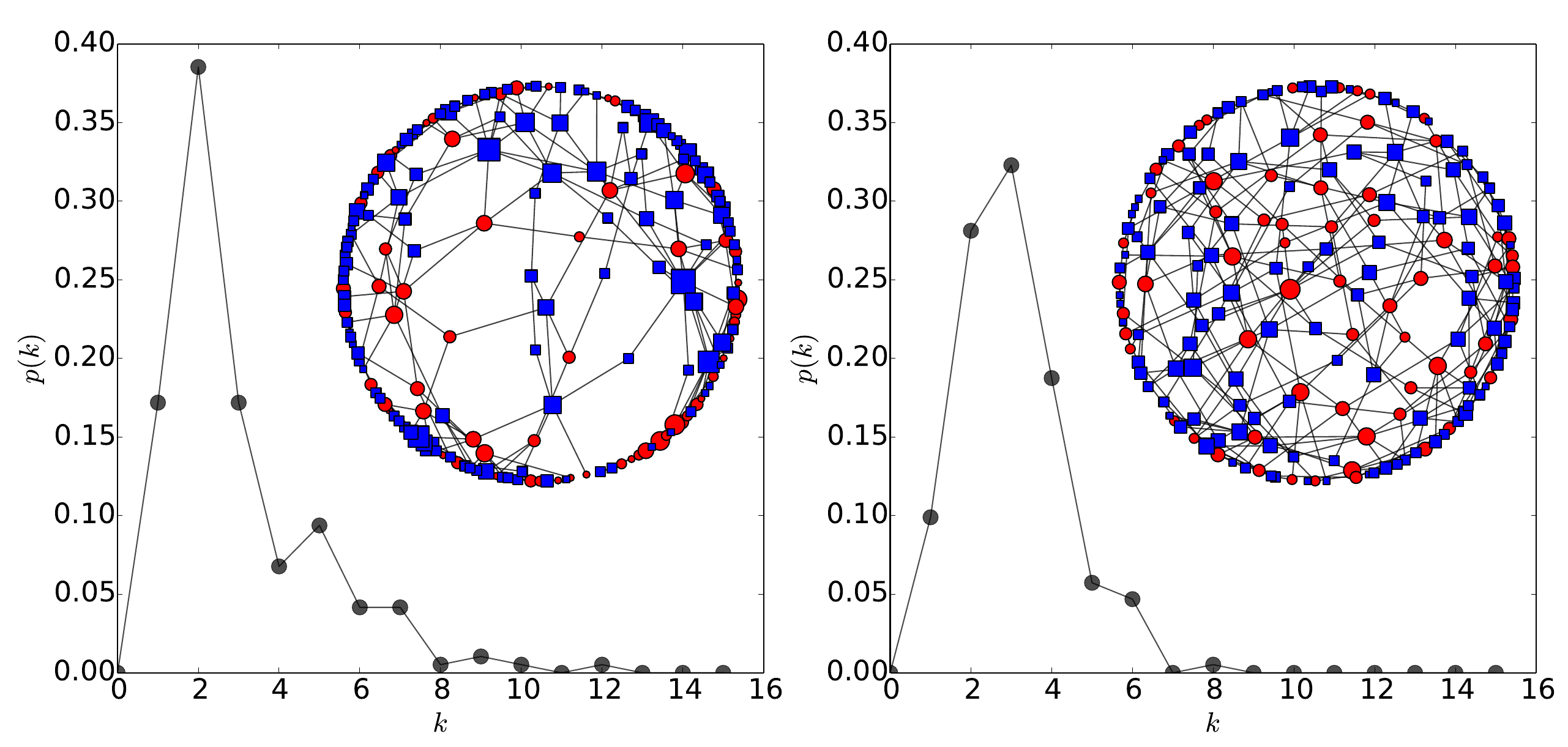}
\caption{An example of an optimized network with  {centralized power generation}. The initial network topology is based in the real Spanish
high voltage
 grid (left handed side) with 128 blue squares and 64 red circles representing consumers and generators, with $P_{i} = -1.0\alpha^2$ and 
 $P_{i} = 2.0\alpha^2$, respectively. The optimized network is shown in the right handed side. For both cases, the vertex's size is proportional to its degree. The corresponding degree distribution for the original and optimized networks are also shown in the panels.
Notice the decreasing of the number of vertices with just one or two neighbors (the so-called  \emph{dead ends}) in the
optimal network.    }
\label{optimization_spanish_grid}
\end{figure}

\begin{figure}[t]
\centering
\includegraphics[scale=0.32]{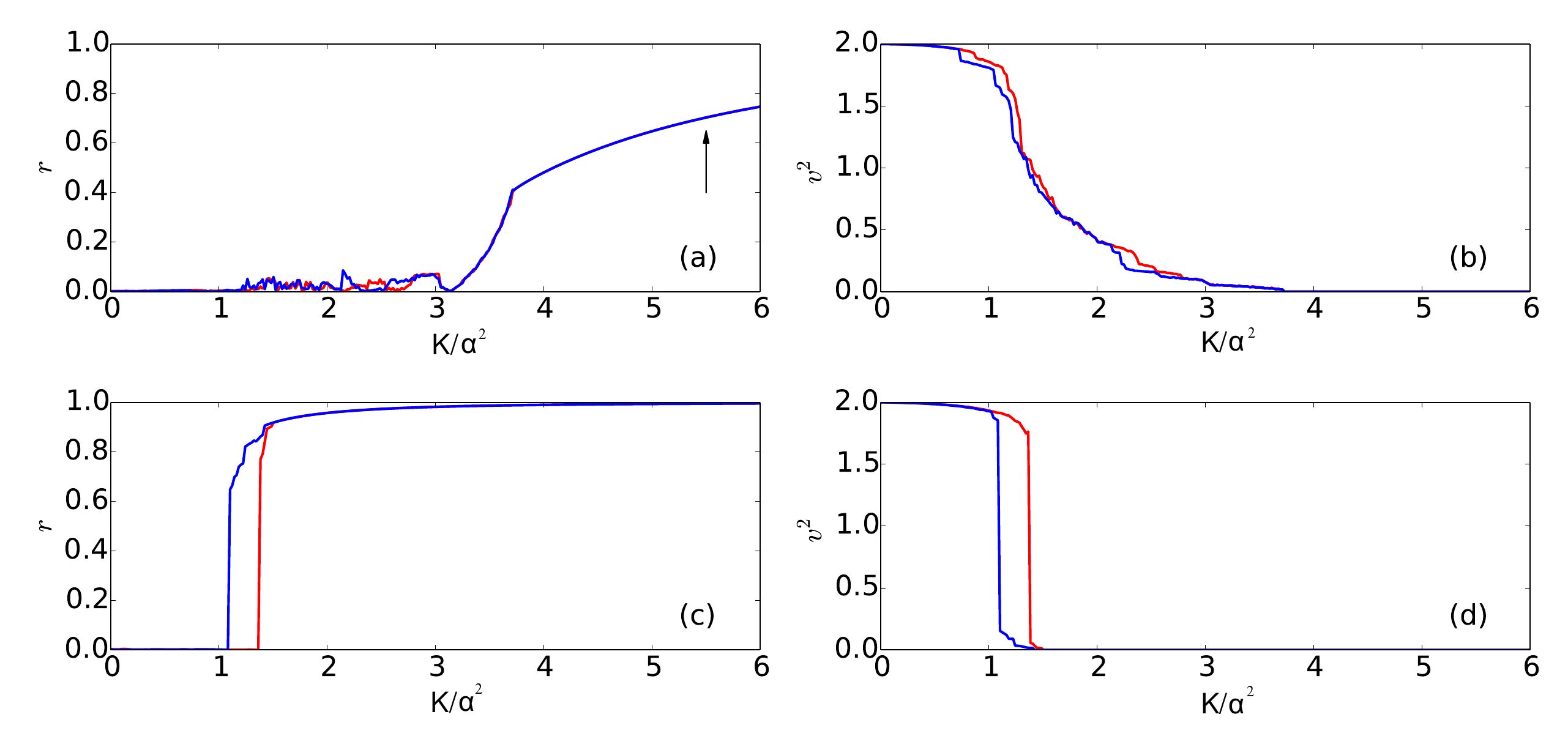}
\caption{Synchronization diagrams for the networks depicted in figure \ref{optimization_spanish_grid}. Both the forward (red) and backward (blue) continuations   are shown. Panels (a) and (b) show the order parameters $r$ and $v^2$  for the non-optimized network. Panels (c) and (d)
correspond to the optimal network and 
 show a considerable improvement in the synchronization properties. The arrow in panel (a) shows where the optimization process was done. Notice the characteristic hysteresis behavior in the synchrony-optimized network.  }
\label{spanish_opt_order_par}
\end{figure}

Having analyzed the decentralized energy production scenarios, we move on to the centralized case, which is perhaps better suited for describing the current status of real power grids. However, for these cases, instead of synthetic networks we will use a 
network with topology based in an approximation of a real-world power grid, namely the Spanish high-voltage grid,
extracted from the approximated western european grid 
 as described by Hutcheon and Bialek in \cite{hutcheon} (See also \cite{Bialek}. The data used here is available at \cite{PowerData}.) Roughly, it is a network of 192 vertices and 287 edges, and its intermediary size  
 is convenient for our analyses since it shows a reasonable compromise between results and required CPU time for the numerical simulations. In the Spanish
network, 64 vertices are generators and the others 128 are consumers. We set the power of each consumers $P = -1\alpha^2$ and, in order to supply the necessary demand, each generator is assumed to have a power $P = 2\alpha^2$. 
Figure \ref{optimization_spanish_grid} shows the original grid in the left handed side and the optimized network   in the right one,
with the corresponding degree distributions. 
An important question in the optimization of real grids is related to the costs of rewiring, since it physically means changing transmission lines, and this is not definitively  an easy task. For the case of figure \ref{optimization_spanish_grid}, we found that the optimized network shares around $37\%$ of edges with the original Spanish grid.
A by-product of the optimization algorithm in the case of centralized power grids is the decrease of the number of vertices with just one or two neighbors, see Figure  \ref{optimization_spanish_grid},
the so-called  \emph{dead ends},
which are known to be extremely vulnerable to cascade faults \cite{menck2014}.
 By reducing the dead ends, our optimization algorithm   also helps to avoid such issues. Our exhaustive numerical experiments show that
the algorithm performs extremely well in optimizing these centralized networks, with a robust improvement in the synchronization properties as 
it is shown in the synchronization diagrams of 
 figure \ref{spanish_opt_order_par}, plotted in the same way as in the centralized case. 

 The behavior of the network topological properties   through the optimization process is very similar to the centralized case.
 For instance, there is a clear trend for the degree distribution standard deviation $\sigma_k$ , which tends to decrease, and
 the fraction $p_{-}$ of edges connecting consumers to generators , which tends to increase during the optimization. As in the
 centralized case, the optimal decentralized  network is typically homogeneous (lower $\sigma_k$) and has many edges connecting consumers to generators 
 (higher $p_{-})$. No appreciable differences in the mean  shortest path length $\langle l \rangle$ and the   clustering coefficient $C$
were observed for the optimal and non-optimal networks. 
We have also considered the robustness of the networks for edge removal by examining  the
  giant component $S(m)$ \cite{zeng2012}  against the same   three  edge removals rules considered in the
  previous section. Our conclusion is the same: our optimization procedure does not weaken the robustness of the original network against edge removals.

\subsection{Impact of consumption peaks}

\begin{figure}[t]
\centering
\includegraphics[scale=0.32]{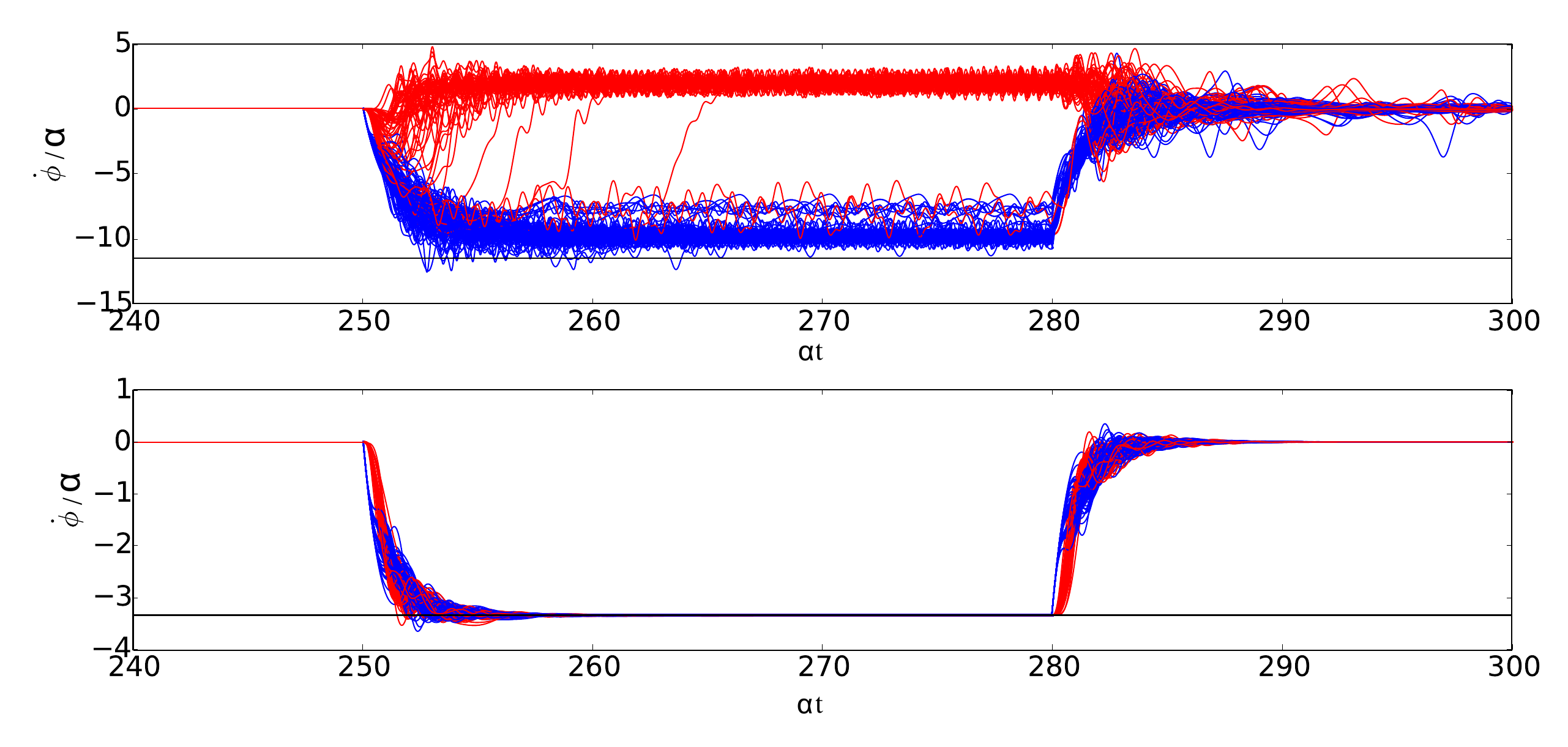}
\caption{An episode of consumption peak for the network on figure \ref{optimization_spanish_grid}, with
$K=4\alpha^2$.  During 30 unities of time $\alpha^{-1}$, 
consumers increase their consumption to $P_i = -6\alpha^2$. Top and bottom panels correspond to the non-optimized and optimized networks, respectively.   The black line is the prediction from equation (\ref{new_sync_state}).}
\label{consume_peak_centralized}
\end{figure}

A consumption peak (or shortage of energy generation) is a sudden event in our network   such that 
\begin{equation}
\label{unb}
  \sum_{i=1}^{N} P_i = \Delta\ne 0,
\end{equation}
typically for a short period of time. This kind of situation 
has been a recurrent topic in many recent works, see \cite{filatrella2008,rohden2012,rohden2014}, for instance, since it mimics
some unbalance events in real-world power grids. We will show the analysis for the centralized case, namely for the
Spanish grid of Fig. \ref{optimization_spanish_grid}, but the decentralized case can be also studied straightforwardly
and conclusions are similar. Notice that
due to the power unbalance Eq. (\ref{triv})  now reads
\begin{equation}
\label{triv1}
\frac{d}{dt} \langle \dot{\phi} \rangle  
  + \alpha  \langle \dot{\phi} \rangle = \Delta,
\end{equation}
from where we see that for  $\Delta \ne 0$, the averaged frequency perturbations does not 
vanish asymptotically, but tends to the value $\Delta/\alpha$, indicating that the synchronized regime, if yet existing,  will have a frequency different
from  $\Omega$, see Eq. (\ref{Omega}). In fact, writing $\phi_{i}(t) = \omega t$ in  (\ref{kuramoto_power_grid_definition}) we
we have that the frequency shift $\omega$ due to the power unbalance  is 
\begin{equation}
\omega = \frac{ \Delta}{\alpha N}  .
\label{new_sync_state}
\end{equation}
The question now is if the power unbalance will allow or not a new synchronized state.  
Figure \ref{consume_peak_centralized} 
shows such situation for both the non-optimal and optimal  networks.
The system is originally balanced, $\Delta = 0$, and suddenly, at $t = 250\alpha^{-1}$ each consumer increases its power to $P = -6\alpha^2$,
until $t = 280\alpha^{-1}$, when they return to the original consumption power $P=-1\alpha^2$. 
For this range of power unbalance, the   non-optimal network cannot attain a synchronized  state 
 during the consumption peak; most of the generators increase their frequencies whereas the consumers decrease. 
On the other hand, for the optimal networks,   both generators and consumers decrease their frequencies at the same pace  
accordingly to (\ref{new_sync_state}), 
in such a way that they reach a new synchronized state with a lower frequency,
which implies a  lower rate of energy dissipation in the whole grid, allowing the supply of the extra  demanded power.
 We stress that this synchronized state with lower frequency, when present, is asymptotically stable for all the parameters used here,
 {\em i.e.}, the dynamics
of the power grid tend to it spontaneously during episodes of consumption peaks.  
The analysis of the case with   $\Delta > 0$, which would correspond to a decrease in consumption,
but keeping the same power pumped into the grid, is completely analogous.

In order to study the range of parameters for which this new synchronized state exists and is indeed stable, 
we consider  the difference between the average frequency of the generators and consumers, $\langle \dot{\phi} \rangle_{g} - \langle \dot{\phi} \rangle_{c}$, during the period of time (that we took to be $30\alpha^{-1}$) in which each consumer has a power increase of $\delta P$. This
situation is depicted in 
\begin{figure}[t]
\centering
\includegraphics[scale=0.32]{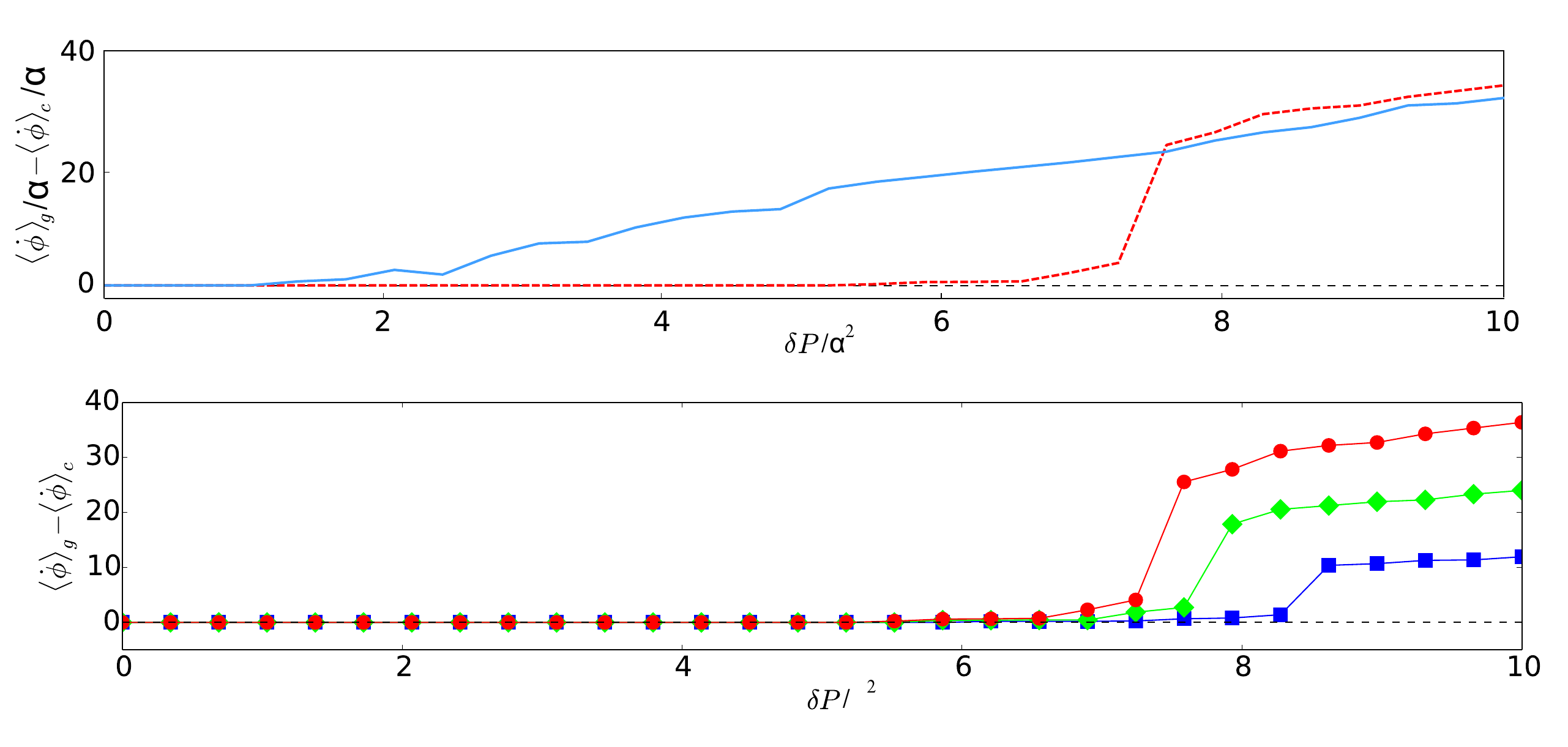}
\caption{Difference in the average frequencies, $\langle \dot{\phi} \rangle_{g} - \langle \dot{\phi} \rangle_{c}$, for generators and consumers, for the non-optimized (blue solid line) and optimized network (red dashed line) of Fig. \ref{optimization_spanish_grid}, with $K=4\alpha^2$,
 during the period $\Delta t=30\alpha^{-1}$ where each consumers increases their  consumption  by $\delta P$, whereas the generators keep the same produced power. }
\label{frequency_average_distuption}
\end{figure}
Figure \ref{frequency_average_distuption}, where we see  that for the non-optimized network, the synchronization state with frequency $\Omega + \omega$ lasts only up to a value of $\delta P \approx 1\alpha^2$. On the other hand, for the optimized network, the average frequencies are the same to a much larger range, until $\delta P \approx 6\alpha^2$.  

\section{Final remarks}

In this paper, we have studied how to optimize the topology of a network of Kuramoto oscillators with inertia  for enhancing its synchronization properties, with special emphasis to the potential  applications in  power grids \cite{filatrella2008}. We introduce a  simple rewiring {hill-climb} 
algorithm and show that it is  possible to
enhance the network synchronization measures and also to 
   reduce the synchronization threshold $K_c$. We applied our proposed algorithm to synthetic random networks and
 also to network inspired in a real world power grid, namely the Spanish high voltage grid. 
The synchrony-optimized power grids obtained by our algorithm have some interesting generic properties besides optimal synchronization patterns.
 For the decentralized generation scenarios, 
the optimized network typically 
  have the majority of edges connecting only consumers to generators. On the other hand, for the case  corresponding to
 centralized generation power grids (the contemporary paradigm), the synchrony-optimized  power grids
 have a minimal number of vertices with just one or two neighbors, known generically as {  dead ends} and which have been
 recently identified as extremely vulnerable and responsible for cascade faults \cite{menck2014}. The optimal
 networks are also  more robust against perturbations mimicking   power supply unbalances as consumption peaks or generation
 shortages. 
 The algorithm can be easily adapted to enhance also the network robustness against edge removals.
 Finally, the synchronization diagram for the optimal networks exhibits  
 a first order phase transition, with a typical hysteresis behavior, while for the non-optimal networks such a transition is typically
 of second order, compare Figs.  \ref{er_opt_er_order_par} and \ref{spanish_opt_order_par}. The implications of this 
dynamical  
 behavior difference  for real-world grids is not yet well understood and certainly deserves a deeper investigation.

 Obviously, the models and 
operations  
 considered here are   extreme simplifications of the real-world power grids.  Nevertheless, 
our results, together with the previous on obtained in \cite{lozano2012,rohden2012,motter2013,witthaut2012,witthaut2013,rohden2014,PNAS,decentral,assort,menck2014},
 show that there is plenty of room for optimizing actual power grids and even
provide
 interesting simple principles to guide the future growth and developments of real-world grids.

\section*{Acknowledgments}

The authors thank CNPq, CAPES and FAPESP (grant 2013/09357-9) for the financial support, and the anonymous referees for
useful comments and criticisms.   
AS thanks Prof. Leon Brenig for several discussions and for 
the warm hospitality at the Free University of Brussels, where the initial part of this work was done.

\end{document}